\documentclass[a4paper,11pt]{article}

\usepackage[left=2.5cm,right=2.5cm,top=2.5cm,bottom=2.5cm]{geometry}
\usepackage{graphicx,amssymb,amsmath,amsthm,mathrsfs,setspace,subcaption,cite,authblk,cancel,bbm,float}
\usepackage[all]{xy}

\usepackage[colorlinks=true,bookmarks=true]{hyperref}

\usepackage{mathtools}

\DeclarePairedDelimiterX\braket[2]{\langle}{\rangle}{#1 \delimsize\vert #2}

\hypersetup{citecolor=blue}

\newcommand{\dif}{\mathrm{d}}
\newcommand{\Eqref}[1]{(\ref{#1})}
\newcommand{\half}{\frac{1}{2}}

\newcommand{\brac}[1]{\left(#1 \right)}
\newcommand{\sbrac}[1]{\left[#1\right]}

\numberwithin{equation}{section}
 
\begin{document}

\title{Energies and angular momenta of periodic Schwarzschild geodesics}

\author[1]{Yen-Kheng Lim\footnote{Email: yenkheng.lim@gmail.com, yenkheng.lim@xmu.edu.my}}

\author[2]{Zhi Cheng Yeo\footnote{Email: thomasyeo4@gmail.com, thomasyeozhicheng@student.usm.my}}

\affil[1]{\normalsize{\textit{Department of Physics, Xiamen University Malaysia, 43900 Sepang, Malaysia}}}

\affil[2]{\normalsize{\textit{School of Physics, Universiti Sains Malaysia, 11800 Gelugor, Malaysia}}}

\date{\normalsize{\today}}
\maketitle
 
\renewcommand\Authands{ and }

\begin{abstract}
  We consider physical parameters of Levin and Perez-Giz's `periodic table of orbits' around the Schwarzschild black hole, where each periodic orbit is classified according to three integers $(z,w,v)$. In particular, we chart its distribution in terms of its angular momenta $L$ and energy $E$. In the $(L,E)$-parameter space, the set of all periodic orbits can be partitioned into domains according to their whirl number $w$, where the limit of infinite $w$ approaches the branch of unstable circular orbits. Within each domain of a given whirl number $w$, the infinite zoom limit $\lim_{z\rightarrow\infty}(z,w,v)$ converges to the common boundary with the adjacent domain of whirl number $w-1$. The distribution of the periodic orbit branches can also be inferred from perturbing stable circular orbits, using the fact that every stable circular orbit is the zero-eccentricity limit of some periodic orbit, or arbitrarily close to one.
\end{abstract}

\section{Introduction} \label{intro}

The study of particle motion in the vicinity of a black hole is often a useful way to probe the gravitational field of the black hole itself. For instance, the motion stars was observed around the galactic black hole in Sgr A* \cite{Ghez:1998ph,Eckart:1996zz,Ghez:2003rt,Genzel:2010zy}. More recently, the direct detection of gravitational waves \cite{LIGOScientific:2016aoc} and the direct imaging \cite{EventHorizonTelescope:2019dse,EventHorizonTelescope:2022wkp} of black hole shadows further makes the case for studying the dynamics of particles in the vicinity of a black hole.

Indeed, it was the problem of gravitational wave detection that was one of the motivations to establish the \emph{periodic table} of black hole orbits by Levin and Perez-Giz \cite{Levin:2008mq}. In their paper, Levin and Perez-Giz argued that any generic bound orbit around a Schwarzschild or Kerr black hole can be seen as arbitrarily close to some periodic, closed orbit. Thus, the set of all bound orbits can be understood through the set of periodic ones. To that end, Levin and Perez-Giz provided a powerful taxonomy scheme where all periodic orbits can be indexed by three non-negative integers, $(z,w,v)$, where $z$ is the \emph{zoom number}, $w$ is the \emph{whirl number}, and $v$ is the \emph{vertex number}.

Subsequently this procedure has been adopted to study particle motion for other spacetimes, such as the Kerr black hole \cite{Levin:2008mq,Levin:2008yp,Perez-Giz:2008ajn,Levin:2009sk,Grossman:2011ps}, Reissner--Nordstr\"{o}m black hole \cite{Misra:2010pu}, the Fisher/Janis--Newman--Winocour spacetime \cite{Babar:2017gsg}, quantum-corrected black holes \cite{Deng:2020yfm}, braneworld black holes \cite{Deng:2020hxw}, black hole surrounded by quintessence \cite{Wang:2022tfo}, Kerr--Sen black holes \cite{Liu:2018vea}, Einstein-\AE ther black holes \cite{Azreg-Ainou:2020bfl}, loop quantum gravity \cite{Tu:2023xab}, among many others. The typical procedure to obtain a periodic orbit is by fixing/choosing a value of angular momentum $L$, then varying the energy $E$ to find a desired orbit. The eccentricity of the orbit will be dependent on the choice of $L$. For a given $L$, some periodic orbits may not exist \cite{Levin:2008mq}. There will be some inherent trial-and-error in finding a periodic orbit of a specified eccentricity.

In this paper, we return to the simplest case of the Schwarzschild spacetime and approach the problem of periodic orbits from a different angle. The new results to be reported in this paper are in Sec.~\ref{sec_periodic} and \ref{sec_LE}, and we briefly summarise it as follows: We give a procedure to calculate precisely the values of $L$ and $E$ required for any choice of $(z,w,v)$ and eccentricity $e$. With this procedure, we can map out the periodic table of orbits in the $(L,E)$-plane. As a result, physical interpretations like the non-relativistic limit (Keplerian ellipses), circular orbits, (non-)existence of certain orbits, and others are made intuitively clear when described in terms of these physical quantities. Various features of periodic orbits derived in \cite{Levin:2008mq} can now be easily read off from the $(L,E)$-plane. 

More precisely, according to the Levin--Perez-Giz scheme, a bound orbit can be parametrised by a dimensionless number $q$. If $q$ is a rational number of the form $q=w+\frac{v}{z}$, for non-negative integers $(z,w,v)$,  we have a periodic orbit which closes after a finite proper time. For a given periodic $q$, there is an additional parameter $0\leq e<1$ which characterises the eccentricity of an orbit. Here, we view $e$ as another independent orbital parameter, so a periodic orbit is uniquely specified by $(q;e)=(z,v,w;e)$. So each periodic orbit $(q;e)$ corresponds to a unique point in $(L,E)$ parameter space. The main goal of this paper is to explore the distribution of periodic orbits in $(L,E)$ parameter space for the Schwarzschild spacetime.

An obvious advantage of using $(L,E)$ to chart the periodic orbits is that they are physical quantities, and are constants of motion for each orbit. Thus its proximity in values to other important orbits like the innermost stable circular orbits (ISCOs), or to the non-relativistic limits, can be seen in the $(L,E)$-plane. In such a diagram, periodic orbits are represented by an infinite sequence of non-intersecting curves of positive slope which converges to the branch of unstable circular orbits. We can parametrise each branch by its eccentricity $e$. The endpoint $e=1$ is where the branch intersects $E=1$, the limit of an escaping trajectory. Running downward in $E$ corresponds to decreasing $e$, until the branch terminates at the branch of stable circular orbits, for which $e=0$. This serves as a graphical realisation of how all stable circular orbits is the zero eccentricity limit of \emph{some} periodic orbit $(z,w,v)$, as was pointed out in \cite{Levin:2008mq}. Furthermore, orbits of different $w$ are subdivisions of domains in the $(L,E)$-plane, and that the $z\rightarrow\infty$ limit for each $w$ is the limit toward the boundary of an adjacent domain $w-1$. In the case $w=0$, the large-$z$ limit is precisely the Kepler ellipse. Additionally, the $w\rightarrow\infty$ limit coincides with the branch of unstable circular orbits. 

The rest of this paper is organised as follows. In Sec.~\ref{sec_eom} we review the geodesic equations for the Schwarzschild spacetime. Readers already familiar with this may skip to Sec.~\ref{sec_periodic}, where we describe our procedure to obtain a periodic orbit of any given $(z,w,v;e)$. In Sec.~\ref{sec_LE} we describe the distribution of periodic orbits in the $(L,E)$ parameter space. The paper concludes in Sec.~\ref{sec_conclusion}. In Appendix \ref{app_taxonomy} we briefly review Levin and Perez-Giz's taxonomy scheme for periodic orbits. As such, Sec.~\ref{sec_eom} and Appendix \ref{app_taxonomy} are reviews of earlier works and can be skipped by the familiar reader. The contents of Sec.~\ref{sec_periodic} are based on the methods of Chandrasekhar's text \cite{ChandrasekharBook}, but modified to the context of determining periodic orbits. Section \ref{subsec_elimits} gives new expressions for the high and low eccentricity periodic orbits. In this sense Sec.~\ref{subsec_procedure} and \ref{subsec_elimits} are ingredients developed uniquely in order to obtain our main results to be reported in this paper, which are in Sec.~\ref{sec_LE}.  

Unless otherwise stated, we work in geometric units where $c=G=1$ and our convention for Lorentzian signature is $(-,+,+,+)$.

\section{Geodesic equations} \label{sec_eom}

The geodesic equations for the Schwarzschild metric are well known and has been covered in detail in most textbooks on GR. (For example, in \cite{ChandrasekharBook}). Here, we review its derivation using the Hamilton--Jacobi approach in Sec.~\ref{subsec_Lagrangian} and obtain the parameters for circular orbits in Sec.~\ref{subsec_circular}. This gives us the opportunity to establish the notation to be used in the rest of the paper. Readers already familiar with Schwarzschild geodesics may perhaps skip ahead to Sec.~\ref{sec_periodic}.

\subsection{Equations of motion} \label{subsec_Lagrangian}

The Schwarzschild metric is
\begin{subequations}\label{metric}
\begin{align}
 \dif s^2&=-f(r)\dif t^2+f(r)^{-1}\dif r^2+r^2\brac{\dif\theta^2+\sin^2\theta\,\dif\phi^2},\\
   f(r)&=1-\frac{2M}{r},
\end{align}
\end{subequations}
where $M$ is the mass of the black hole. In this paper, we are only interested in the region exterior to the black hole, that is, $r>2M$. Trajectories of time-like particles will be curves parametrised by its proper time, $x^\mu(\tau)=\brac{t(\tau),r(\tau),\theta(\tau),\phi(\tau)}$. For a test particle, the Lagrangian for geodesic motion is $\mathcal{L}(x,\dot{x})=\half g_{\mu\nu}\dot{x}^\mu\dot{x}^\nu$, where over-dots denote derivatives with respect to proper time $\tau$. In the Schwarzschild spacetime, it is explicitly
\begin{align}
 \mathcal{L}(x,\dot{x})&=\half \brac{-f\dot{t}^2+\frac{\dot{r}^2}{f}+r^2\dot{\theta}^2+r^2\sin^2\theta\dot{\phi}^2}.
\end{align}
The canonical momenta are $p_\mu=\frac{\partial\mathcal{L}}{\partial\dot{x}^\mu}$, which gives
\begin{align}
 p_t&=-f\dot{t},\quad p_r=\frac{\dot{r}}{f},\quad p_\theta=r^2\dot{\theta},\quad p_\phi=r^2\sin^2\theta\,\dot{\phi}.\label{canonical_p}
\end{align}
After taking the Legendre transform to obtain the Hamiltonian, the Hamilton--Jacobi equation is
\begin{align}
 \half\sbrac{-\frac{1}{f}\brac{\frac{\partial S}{\partial t}}^2+f\brac{\frac{\partial S}{\partial r}}^2+\frac{1}{r^2}\brac{\frac{\partial S}{\partial\theta}}^2+\frac{1}{r^2\sin^2\theta}\brac{\frac{\partial S}{\partial\phi}}^2}+\frac{\partial S}{\partial\tau}=0.
\end{align}
Using the standard methods, the Hamilton--Jacobi equation is completely separated, which results in the equations of motion
\begin{subequations}\label{Sch_eom}
\begin{align}
 \dot{t}&=-\frac{E}{f},\quad\theta=\frac{\pi}{2},\label{dot_t}\\
  \dot{\phi}&=\frac{L}{r^2},\label{dot_phi}\\
 \dot{r}^2&=E^2-U_{\mathrm{eff}},\quad U_{\mathrm{eff}}=\brac{1+\frac{L^2}{r^2}}\brac{1-\frac{2M}{r}}, \label{Ueff}
\end{align}
\end{subequations}
where due to spherical symmetry, we have taken $\theta=\frac{\pi}{2}=\mathrm{constant}$ without loss of generality. This also fixes the value of the separation constant in terms of $L$. Here, $E$ and $L$ are the particle's conserved energy and angular momentum, respectively. Equation \Eqref{Sch_eom} are sufficient to determine all time-like geodesics of interest in this paper. Nevertheless, it is convenient to have at hand a second order equation for $r$,
\begin{align}
 \ddot{r}&=\frac{f'\dot{r}^2}{2f}-\frac{f'E^2}{2f}+\frac{L^2f}{r^3}, \label{ddot_r}
\end{align}
which is obtained by applying the Euler--Lagrange equation.

A differential equation between $r$ and $\phi$ can be obtained by taking $\frac{\dot{r}}{\dot{\phi}}=\frac{\dif r}{\dif\phi}$, giving
\begin{align}
 \frac{\dif r}{\dif\phi}&=\pm r^2\sqrt{\frac{E^2-1}{L^2}+\frac{2M}{L^2r}-\frac{1}{r^2}+\frac{2M}{r^3}}. \label{drdphi}
\end{align}

\subsection{Circular orbits and the \texorpdfstring{$(L,E)$}{(L,E)} parameter space} \label{subsec_circular}

As we will see later, all branches of periodic orbits are continuously connected to the branch of stable circular orbits. As such let us review the parameters describing circular orbits in Schwarzschild spacetime.

\begin{figure}
 \centering
 \includegraphics{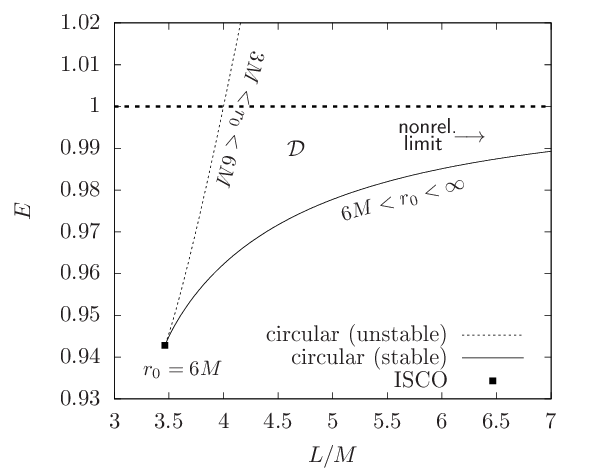}
 \caption{Values of angular momenta $L$ and energy $E$ for circular orbits in Schwarzschild spacetime. The domain $\mathcal{D}$ is the region bounded by the two branches of circular orbits and the line $E=1$. The non-relativistic, or Keplerian limit is where $L/M\rightarrow\infty$ and $E\rightarrow1$.}
 \label{fig_circular}
\end{figure}

We define circular orbits as orbits of constant $r=r_0$. From Eq.~\Eqref{Ueff}, this is attained by
\begin{align}
 E^2=U_{\mathrm{eff}},\quad U_{\mathrm{eff}}'=0.
\end{align}
Solving for $E$ and $L$, we get
\begin{align}
 E=\frac{r_0-2M}{\sqrt{r_0(r_0-3M)}},\quad L=r_0\sqrt{\frac{M}{r_0-3M}}, \label{circular_EL}
\end{align}
corresponding to a circular orbit of radius $r=r_0$. To check the stability of circular orbits we turn to the second derivative of $U_{\mathrm{eff}}$ evaluated at the circular orbits. We find
\begin{align}
 \left.U''_{\mathrm{eff}}\right|_{\mathrm{circular}}=\frac{2M(r_0-6M)}{r_0^3(r_0-3M)}. \label{Upp}
\end{align}
Therefore we see that $U_{\mathrm{eff}}''>0$ for $r_0>6M$, corresponding to the branch of stable circular orbits, while $U_{\mathrm{eff}}''<0$ for $3M<r_0<6M$ for the branch of unstable circular orbits. No circular orbits exist for $r<3M$. Following the standard literature, the critical point $r_0=6M$ is called the \emph{innermost stable circular orbit} (ISCO). The circular orbits (and subsequently, periodic orbits in the later sections) can be mapped on a $(L,E)$-parameter space, as shown in Fig.~\ref{fig_circular}. The dotted and solid curves represent the branch of stable and unstable circular orbits, respectively. It can be viewed as a parametrised curve according to Eq.~\Eqref{circular_EL}, parametrised by the circle radius $r_0$.

Let us denote by $\mathcal{D}$ the set 
\begin{align}
 \mathcal{D}=\left\{\mbox{values of $(L,E)$ of bound, non-plunging orbits} \right\}
\end{align}
In the $(L,E)$ plane, $\mathcal{D}$ is the region bounded by the curves for circular orbits and the line $E=1$. (See Fig.~\ref{fig_circular}.) In this region, the equation $E^2=U_{\mathrm{eff}}$ has three distinct real roots in $r$, and there exist a domain of $r$ between two of the roots for which $E^2>U_{\mathrm{eff}}$. Hence the particle can be bound in radii between these two roots without falling into the horizon. These are the set of bound, non-plunging orbits and the subject of main interest in this paper. All periodic orbits will have angular momenta and energies $(L,E)$ contained in this region.
It is also worth recalling where in the $(L,E)$-space does the non-relativistic (i.e., Keplerian) regime lie. Restoring to standard units,
\begin{align}
 E\rightarrow\frac{\mathcal{E}}{mc^2}+1,\quad L\rightarrow\frac{J}{mc},\quad M\rightarrow\frac{G\mathcal{M}}{c^2},
\end{align}
where $\mathcal{E}$, $J$, and $m$ are the test particle's energy, angular momentum, and mass in standard units. $\mathcal{M}$ is the black hole mass, and $c$ and $G$ are the speed of light and gravitational constant, respectively. To recover the non-relativistic limit, we take 
\begin{align*}
 E^2=\brac{\frac{\mathcal{E}}{mc^2}+1}^2&\simeq 1+\frac{2\mathcal{E}}{mc^2}+\mathcal{O}\brac{\brac{\textstyle{\frac{\mathcal{E}}{mc^2}}}^2},\\
 \frac{G\mathcal{M}}{c^2r}&\ll 1.
\end{align*}
This latter condition occurs at large $r$ far away from the black hole. In this limit, Eq.~\Eqref{drdphi} reduces to 
\begin{align*}
 \frac{\dif r}{\dif\phi}=\pm\sqrt{\frac{2m\mathcal{E}}{J^2}-\frac{1}{r^2}+\frac{2G\mathcal{M}m^2}{J^2r}},
\end{align*}
which is precisely the Keplerian equation of motion, for which the exact solutions are ellipses. In terms to our $(L,E)$ plane, which is in geometric units, the non-relativistic limit is where $E\rightarrow 1$ and $L/M=Jc/G\mathcal{M}\rightarrow\infty$, as indicated by `\textsf{nonrel.~limit}' in Fig.~\ref{fig_circular}.

\section{Periodic orbits and their eccentricities} \label{sec_periodic}

In Ref.~\cite{Levin:2008mq}, Levin and Perez-Giz gave a powerful classification scheme where a triplet of integers $(z,w,v)$ is associated to each periodic orbit. In the present paper, we are mainly interested in the set of possible angular momenta $L$ and energies $E$ for an orbit of given $(z,w,v)$. Here we review the procedure to obtain periodic orbits, but modifying it to place some emphasis on the eccentricity and latus rectum of the orbits. This will be useful to establish a map of periodic orbits in the $(L,E)$-plane.

\subsection{Analytical solution} \label{subsec_anal}

Introducing the substitution $u=1/r$, Eq.~\Eqref{drdphi} becomes
\begin{align}
 \frac{\dif u}{\dif\phi}=\pm\sqrt{P(u)},\quad P(u)=-\frac{1-E^2}{L^2}+\frac{2M}{L^2}u-u^2+2Mu^3. \label{Sch_Pu}
\end{align}
Clearly, the physically allowable domain for $u$ is such that $P(u)\geq 0$. In particular, the roots of $P$ specify the boundary of these domains and represent the turning point in the $r$-motion, as $\dot{r}=0$ at these points. We restrict our attention to parameters such that $P(u)$ has three real roots, ordered by
\begin{align}
 u_1\leq u_2\leq u_3.
\end{align}
Since the leading coefficient of $P$ is positive, $P(u)$ is non-negative for $u_1\leq u\leq u_2$ and $u\geq u_3$. In this paper, we are only interested in bound orbits which do not fall into the black-hole horizon. Hence we seek trajectories lying in a finite domain, which is
\begin{align}
 u_1\leq u\leq u_2.
\end{align}
In other words, $r_1=1/u_1$ corresponds to the aphelion and $r_2=1/u_2$ corresponds to the perihelion, and the trajectory oscillates between these two radii. The largest root $u_3$ goes to infinity in the non-relativistic limit where $P(u)$ becomes quadratic.

In terms of these roots, the polynomial $P$ can be we rewritten such that
\begin{align}
 \frac{\dif u}{\dif\phi}&=\pm\sqrt{P(u)}=\pm\sqrt{2M(u_3-u)(u_2-u)(u-u_1)}.
\end{align}
We choose initial conditions $u=u_1$ at $\phi=0$. With $L>0$, this means the subsequent motion is $u$ increasing while $\phi$ increases and implies that $\frac{\dif u}{\dif\phi}>0$ as $u$ evolves from $u_1$ toward $u_2$. For this domain we take the positive branch of the square root. Then Eq.~\Eqref{Sch_Pu} can be integrated to give
\begin{align*}
 \frac{1}{\sqrt{2M}}\int_{u_1}^u\frac{\dif u'}{\sqrt{(u_3-u')(u_2-u')(u'-u_1)}}=\phi.
\end{align*}
This integral is evaluated exactly as\footnote{See, for instance, \cite{gradshteyn2014table} (pg 254, 3.131-3).}
\begin{align*}
 \phi(u)&=\frac{2}{\sqrt{2M(u_3-u_1)}}\mathrm{F}\brac{\arcsin\sqrt{\frac{u-u_1}{u_2-u_1}},\sqrt{\frac{u_2-u_1}{u_3-u_1}}}, \label{phi_u}
\end{align*}
where $\mathrm{F}\brac{x,k}$ is the incomplete elliptic integral of the first kind. This can be inverted to give
\begin{align}
 u(\phi)=u_1+(u_2-u_1)\mathrm{sn}^2\brac{\sqrt{2M(u_3-u_1)}\,\textstyle{\frac{\phi}{2}},\;\textstyle{\sqrt{\frac{u_2-u_1}{u_3-u_1}}}},
\end{align}
where $\mathrm{sn}\brac{\theta,k}$ is the Jacobi elliptic sine function. Restoring $r=1/u$, the exact solution for the trajectory is
\begin{align}
 r(\phi)=\frac{1}{u_1+(u_2-u_1)\mathrm{sn}^2\brac{\sqrt{2M(u_3-u_1)}\,\textstyle{\frac{\phi}{2}},\;\textstyle{\sqrt{\frac{u_2-u_1}{u_3-u_1}}}}}. \label{sch_anal}
\end{align}

\subsection{Eccentricity and latus rectum}

Following Chandrasekhar \cite{ChandrasekharBook} (page 103, Eq.~(114)), we parametrise the roots with
\begin{align}
 u_3=\frac{1}{2M}-\frac{2}{\lambda},\quad u_2=\frac{1+e}{\lambda},\quad u_1=\frac{1-e}{\lambda}, \label{parametrize_roots}
\end{align}
where $e$ is the \emph{eccentricity} and $\lambda$ is the \emph{latus rectum}. Note that $e=\frac{u_2-u_1}{u_2+u_1}$, which is the same definition of eccentricity used in \cite{Levin:2008mq}.  By comparing coefficients with \Eqref{Sch_Pu}, they are related to $E$, $L$, and $M$ by
\begin{align}
 \frac{1-E^2}{L^2}=\frac{(\lambda-4M)(1-e^2)}{\lambda^3},\quad \frac{M}{L^2}=\frac{\lambda-3M-Me^2}{\lambda^2}. \label{EL_to_el}
\end{align}
This parametrisation also leads to
\begin{align}
 u_3-u_1&=\frac{1}{2M}+\frac{e-3}{\lambda},\quad u_2-u_1=\frac{2e}{\lambda},
\end{align}
so that the exact solution \Eqref{sch_anal} can be expressed in terms of the orbital parameters as
\begin{align}
 r(\phi)&=\frac{\lambda}{1-e+2e\mathrm{sn}^2\brac{\sqrt{1+\frac{2M(e-3)}{\lambda}}\;\frac{\phi}{2},\sqrt{\frac{4Me}{\lambda+2M(e-3)}}}}.
\end{align} 
This completes the exact solution to the equations of motion.

However, to seek periodic orbits, we shall use Eq.~\Eqref{phi_u} instead where $\phi$ is treated as a function of $u=1/r$. During the trajectory, the radial coordinate of the particle oscillates between its maximum and minimum values $1/u_2\leq r\leq 1/u_1$. During the time it executes one period of this oscillation, the evolution of the azimuthal angle is 
\begin{align}
 \Delta\phi_r&=2\phi(u_2)=\frac{2}{\sqrt{2M(u_3-u_1)}}K\brac{\textstyle{\sqrt{\frac{u_2-u_1}{u_3-u_1}}}}\nonumber\\
 &=\frac{4}{\sqrt{1+\frac{2M}{\lambda}(e-3)}}K\brac{\textstyle{\sqrt{\frac{4Me}{\lambda+2M(e-3)}}}},
\end{align}
where $K(k)$ is the complete elliptic integral of the first kind with elliptic modulus $k$. Furthermore, a parameter $q$ can be defined by \cite{Levin:2008mq}
\begin{align}
 q+1=\frac{\Delta\phi_r}{2\pi}=\frac{2}{\pi\sqrt{1+\frac{2M}{\lambda}(e-3)}}K\brac{\textstyle{\sqrt{\frac{4Me}{\lambda+2M(e-3)}}}}. \label{qplus1}
\end{align}
A periodic orbit occurs if $\Delta\phi_r$ is a rational multiple of $2\pi$, which means that $q$ is a rational number. More precisely, this is when $q=w+\frac{v}{z}$ for three non-negative integers $(z,w,v)$ where $z$, $w$, and $w$ are the \emph{zoom}, \emph{whirl}, and \emph{vertex} numbers, respectively, according to the taxonomy scheme of Levin and Perez-Giz \cite{Levin:2008mq}. 

We recall that the zoom number $z$ gives the number of `leaves', or `petals' of the periodic orbit. The vertex number $v$ determines order of how each petal is traced out for a given $z$. The integer $v$ must be relatively prime to $z$. Furthermore in the case $z=1$, there is only one petal and $v$ is defined to be 0 in this case. Therefore $v$ and $z$ have the following relationship: 
\begin{align}
 1&\leq v\leq z-1,\quad \mbox{if}\quad\mbox{$z$ and $v$ are co-prime},\nonumber\\
 v&=0,\quad \mbox{if}\quad z=1. \label{zv_constraint}
\end{align}
Finally, the whirl number $w$ tells us the number of laps it executes around the black hole in the time between successive petals. Further details about the basic notions of the Levin--Perez-Giz taxonomy is reviewed in Appendix \ref{app_taxonomy}.

\subsection{Obtaining a periodic orbit of a desired \texorpdfstring{$(z,v,w)$}{(z,v,w)} and \texorpdfstring{$e$}{e}} \label{subsec_procedure}

From Eq.~\Eqref{qplus1}, we see that periodic orbits occur if $K\brac{\sqrt{\frac{4Me}{\lambda+2M(e-3)}}}$ is a rational multiple of $\brac{\pi\sqrt{1+\frac{2M}{\lambda}(e-3)}}^{-1}$. Since $K(k)$ is a monotonic function of its argument,
we can typically find a $\lambda$ for a desired periodic orbit $(z,w,v)$ of eccentricity $e$. For each pair $(e,\lambda)$, its corresponding energy and angular momentum found using Eq.~\Eqref{EL_to_el} to give
\begin{align}
 E=\sqrt{\frac{\lambda^2-4M\lambda+4M^2-4M^2e^2}{\lambda(\lambda-3M-Me^2)}},\quad L=\lambda\sqrt{\frac{M}{\lambda-3M-Me^2}}.\label{solve_EL}
\end{align}

Therefore, a practical recipe for obtaining a desired periodic orbit is as follows: Choose a periodic orbit $(z,w,v)$ and its eccentricity $e$, then $q=w+\frac{v}{z}$. With the required $q$ and $e$ at hand, use Eq.~\Eqref{qplus1} to solve for $\lambda$. The corresponding physical parameters $E$ and $L$ which creates the orbit are then given by the formulas in \Eqref{solve_EL}. Only the determination of $\lambda$ requires numerical root-finding.

\begin{figure}
  \begin{subfigure}[b]{0.33\textwidth}
    \centering
    \includegraphics[scale=0.75]{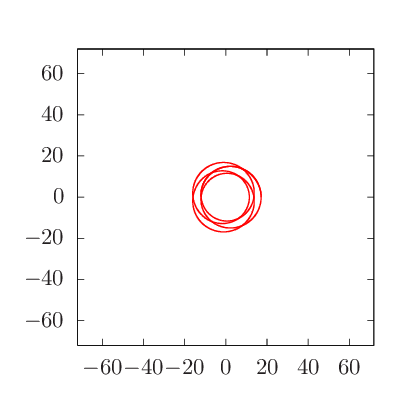}
    \caption{$e=0.2$.}
    \label{fig_301_traj_e_0.2}
  \end{subfigure}
  \begin{subfigure}[b]{0.33\textwidth}
    \centering
    \includegraphics[scale=0.75]{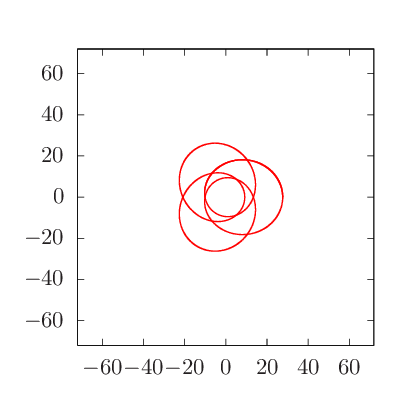}
    \caption{$e=0.5$.}
    \label{fig_301_traj_e_0.5}
  \end{subfigure}
  \begin{subfigure}[b]{0.33\textwidth}
    \centering
    \includegraphics[scale=0.75]{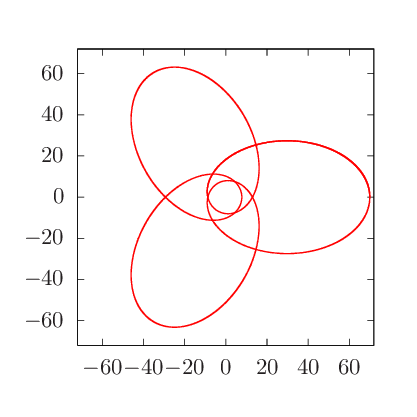}
    \caption{$e=0.8$.}
    \label{fig_301_traj_e_0.8}
  \end{subfigure}
  \caption{The $(3,0,1)$ orbits for different eccentricities.}
  \label{fig_301_traj}
\end{figure}

As an example, let us obtain the orbit $(3,0,1)$ for three choices of eccentricities, $e=0.2$, $e=0.5$ and $e=0.8$. For this orbit, $q=1$. Solving Eq.~\Eqref{qplus1} gives $\lambda=13.73203531M$ for $e=0.2$, $\lambda=13.82431376M$ for $e=0.5$, and $\lambda=13.99183759M$ for $e=0.8$. Next, using Eq.~\Eqref{solve_EL} gives the corresponding energies and angular momenta:
\begin{align}
 e=0.2:\quad E&=0.9676607923,\quad L=4.199568983M,\nonumber\\
 e=0.5:\quad E&=0.9744720033,\quad L=4.251258230M,\nonumber\\
 e=0.8:\quad E&=0.9875046762,\quad L=4.348765909M. \label{orbit_301_vals}
\end{align}
The orbits for these three choices of $e$'s are plotted in Cartesian coordinates in Figs.~\ref{fig_301_traj_e_0.2}, \ref{fig_301_traj_e_0.5}, and \ref{fig_301_traj_e_0.8}.

\subsection{High and low eccentricity limits} \label{subsec_elimits}

\textbf{The limit $e\rightarrow0$.} Given any periodic orbit, the limit $e\rightarrow0$ reduces to that of a stable circular orbit. In fact, this was the prescription given by Levin and Perez-Giz to assign a $(z,w,v)$ triplet for each circular orbit \cite{Levin:2008mq}. Our present task here is to obtain and explore an explicit formula based on this prescription. With this formula one can determine the radius $r_0$ of the resulting circular-orbit limit of any given $q=w+\frac{v}{z}$. As will be seen in the following section, this will be useful in understanding the distribution of periodic orbits in $\mathcal{D}$. 

Let a circular orbit be characterised by its radius $r_0$. Its energy and angular momentum are as given by Eq.~\Eqref{circular_EL}. We then introduce a small perturbation by writing
\begin{align}
 r(\tau)&=r_0+\varepsilon r_1(\tau).\label{r_perturb}
\end{align}
Substituting \Eqref{r_perturb} into Eq.~\Eqref{ddot_r} and keeping up to first order in $\varepsilon$, we find\footnote{Equivalently, $\left.U_{\mathrm{eff}}''\right|_{\mathrm{circular}}=\frac{2}{r_0}\Omega^2$, as can be seen by comparing with Eq.~\Eqref{Upp}.}
\begin{align*}
 \ddot{r}_1&=-\Omega^2 r_1,\quad \Omega=\frac{1}{r_0}\sqrt{\frac{M(r_0-6M)}{r_0(r_0-3M)}}.
\end{align*}
Suppose we choose initial conditions such that $r_1(\tau)=A\cos\Omega\tau$, where $A\sim\mathcal{O}(1)$. Putting this solution into Eq.~\Eqref{dot_phi} leads to
\begin{align}
 \phi(\tau)&=\frac{L_0}{r_0^2}\brac{\tau-\frac{2\varepsilon A}{\Omega r_0}\sin\Omega\tau+\ldots}.
\end{align}
Therefore, the increment in $\phi$ as the particle returns to its starting radius is when $\tau=\frac{2\pi}{\Omega}$, or, to leading order,
\begin{align}
 \frac{\Delta\phi_r}{2\pi}&\simeq\frac{L_0}{r_0^2\Omega}=\sqrt{\frac{r_0}{r_0-6M}}.
\end{align}
In other words, this is an approximate formula for $q+1$ for perturbed circular orbits. If $q$ is a rational number, we then have periodic orbits for low eccentricities:
\begin{align}
 q+1=w+\frac{v}{z}+1=\sqrt{\frac{r_0}{r_0-6M}},\quad 0\lesssim e\ll 1.\label{perturb_q}
\end{align}
This furnishes an explicit formula for the prescription of a $(z,w,v)$-triplet for a circular orbit according to \cite{Levin:2008mq}. Since stable circular orbits are characterised by $r_0>6M$, then the set of \emph{all} stable circular orbits can be mapped to the positive real line $\mathbb{R}_{+}$. Just as every point along $\mathbb{R}_{+}$ is arbitrarily close to some rational number, for every stable circular orbit of radius $r_0$, there exist a periodic orbit whose $e\rightarrow0$ limit is arbitrarily close to it.

\textbf{The limit $e\rightarrow1$.} Unlike the zero eccentricity limit, the case $e\rightarrow1$ does not admit a nice perturbative expression for a given $q$ analogous to Eq.~\Eqref{perturb_q}. However, we can instead consider the limit $e\rightarrow1$ for some fixed $L$. From Eq.~\Eqref{solve_EL}, at fixed $L$, the latus rectum is
\begin{align}
 \lambda_\pm=\frac{L\brac{L\pm\sqrt{L^2-16M^2}}}{2M}.
\end{align}
Using this expression for $\lambda$, one can use Eq.~\Eqref{qplus1} to check the value of $q$. We find that $\lambda_-$ leads to complex values of $q$, so we discard it. Taking $\lambda_+$ and assuming $L\geq4M$, this gives
\begin{align}
 q_{\mathrm{max}}=-1+\frac{2}{\pi}\sqrt{\frac{L\brac{L+\sqrt{L^2-16M^2}}}{L^2-8M^2+L\sqrt{L^2-16M^2}}}K\brac{2M\sqrt{\frac{2}{L^2-8M^2+\sqrt{L^2-16M^2}}}}.\label{q_max}
\end{align}
We denote by $q_{\mathrm{max}}$ as the orbit $q$ with $e\rightarrow1$ at a given $L$. Thus, Eq.~\Eqref{q_max} gives an explicit formula for $q_{\mathrm{max}}$ which appears in Eq.~(12) of \cite{Levin:2008mq}.

\section{Distribution of periodic Schwarzschild orbits in \texorpdfstring{$(L,E)$}{(L,E)} parameter space} \label{sec_LE}

With the tools developed and reviewed in the previous sections, our task in this section is to establish a map of periodic orbits in $(L,E)$-parameter space. This involves understanding how the energy and angular momenta of periodic orbits are distributed in the domain $\mathcal{D}$ in Fig.~\ref{fig_circular}.

\subsection{Preliminary observations}

\begin{figure}
 \centering
    \includegraphics[scale=1]{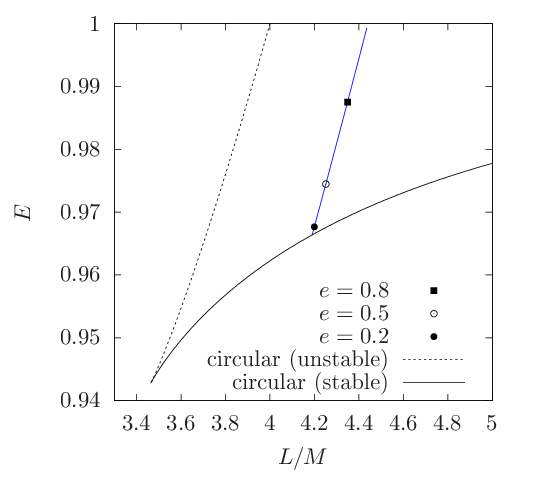}
    \caption{The blue solid curve in shows the values of angular momentum and energies for the orbit $(3,0,1)$; orbits with $e\rightarrow 1$ approaches $E=1$ and orbits with $e\rightarrow0$ approaches the dotted black curve corresponding to circular orbits.}
    \label{fig_301_param}
\end{figure}

As a starting point, consider the concrete example of the orbit $(3,0,1)$. In the previous section, we have explicitly calculated the energies and angular momenta of the cases $e=0.2$, $0.5$, and $0.8$ in Eq.~\Eqref{orbit_301_vals}. Their values are marked in $(L,E)$-space shown in Fig.~\ref{fig_301_param}. Now, obtaining the corresponding $L$ and $E$ for all eccentricities $0<e<1$ gives all the points along the solid blue curve in Fig.~\ref{fig_301_param}. In the following, let us refer to the set all orbits of a given $q$ ---whether it is rational or irrational---as a \emph{q-branch}, or simply a \emph{branch}.\footnote{We choose the word \emph{branch} because from Fig.~\ref{fig_301_param} it is a curve that branches off from the curve of stable circular orbits.} For periodic orbits, $q$ is rational and we call the set a \emph{periodic orbit branch}.

This example of the $(3,0,1)$-branch makes clear the following observations:
\begin{enumerate}
 \item The branch all orbits with taxonomy $(3,0,1)$ lie along a curve that starts from the curve of stable circular orbits and terminates at $E=1$.
 \item The branch meets the branch of stable circular orbits in the limit $e\rightarrow0$. In the small neighbourhood of $e\simeq0$, the orbit is a perturbation of a stable circular orbit. So the values of $L$ and $E$ at the intersection point can be found with Eq.~\Eqref{circular_EL}, where $r_0$ is determined from $(z,w,v)$ using Eq.~\Eqref{r0_zwv}. 
 \item In the other direction, the branch meets the line $E=1$ in the limit $e\rightarrow1$.
\end{enumerate}

While the above discussion was for the explicit case $(3,0,1)$, the above statements also hold for any generic $q=w+\frac{v}{z}$. In particular, each periodic orbit branch is a curve lying in domain $\mathcal{D}$. It is a one-dimensional curve parametrised by eccentricity $e$. The endpoint $e\rightarrow0$ is the zero-eccentricity limit where the branch meets the curve of stable circular orbits at a point unique to each $(z,w,v)$. The specific point $(L,E)$ is determined from \Eqref{circular_EL} with $r_0$ determined from Eq.~\Eqref{perturb_q}.

In other words, each periodic orbit branch emanates from the curve of stable circular orbits at different points. That is, there is no single point on the stable circular orbit curve where two or more periodic branches emanate from. Conversely, no distinct periodic branches have the same zero-eccentricity limit leading to circular orbits of the same radius. 

Furthermore, these periodic orbit branches do not intersect each other anywhere else in $\mathcal{D}$. If periodic orbit branches intersect, the intersection point implies a value of $L$ and $E$ whose orbit has multiple distinct taxonomies, which is impossible. 

All orbits have the extreme eccentricity limit $e\rightarrow1$, which is the limit to an escaping trajectory. Therefore all the periodic orbit branches meet the line $E=1$, also at unique points due to the arguments of the previous paragraph.  

From these discussions, we conclude that in the $(L,E)$-plane, the set of all periodic orbits consists of a family of non-intersecting curves, each emanating from unique points on the curve of stable circular orbits, and terminating at the line $E=1$. Because the branches do not intersect, the distribution and ordering of the branches can be understood through its distribution and ordering of points of emanation from the curve of stable circular orbits. These points can be easily obtained using Eq.~\Eqref{perturb_q}. We shall turn to this task in the following subsection.

\subsection{Parametrisation of periodic orbits along stable circular orbits} \label{subsec_stableparam}

In $(L,E)$-space, the branches of periodic orbits are all represented by non-intersecting curves which connects the branch of stable circular orbits to the curve $E=1$. All the curves do not intersect each other, so the ordering among periodic orbits is preserved in the domain $\mathcal{D}$ for any eccentricity $e$. This means that it suffices to study the sequence of periodic orbits in the limit $e\rightarrow0$. In this limit, the radius of the resulting circular orbits can be found by solving Eq.~\Eqref{perturb_q} for the radius, giving
\begin{align}
 r_{0q}=r_{0}(z,w,v)=\frac{6M(q+1)^2}{q(q+2)}=\frac{6M(z+v+wz)^2}{(wz+v)(wz+v+2z)}.\label{r0_zwv}
\end{align}
Since for periodic orbits $q=w+\frac{v}{z}$, we use the notations $r_{0q}$ and $r_{0}(z,w,v)$ interchangeably. In particular the latter will be useful when viewing the sequence orbits increasing $z$, but fixed $w$ and $v$, for example. In any case Eq.~\Eqref{r0_zwv} gives the radius of the zero-eccentricity limit of a periodic orbit of a given $q=w+\frac{v}{z}$. Using the $r_{0q}$ calculated from a given $q$, the energy and angular momenta (hence, the point along the stable circular orbit curve) is determined using Eq.~\Eqref{circular_EL}.

Using Eq.~\Eqref{r0_zwv}, one can see the precise sequence of branches in the $(L,E)$ space based on its intersection point with the stable circular orbit branch. Hence the description of the branches can be put on a more precise footing.

\begin{figure}
 \centering 
 \includegraphics[width=0.8\textwidth]{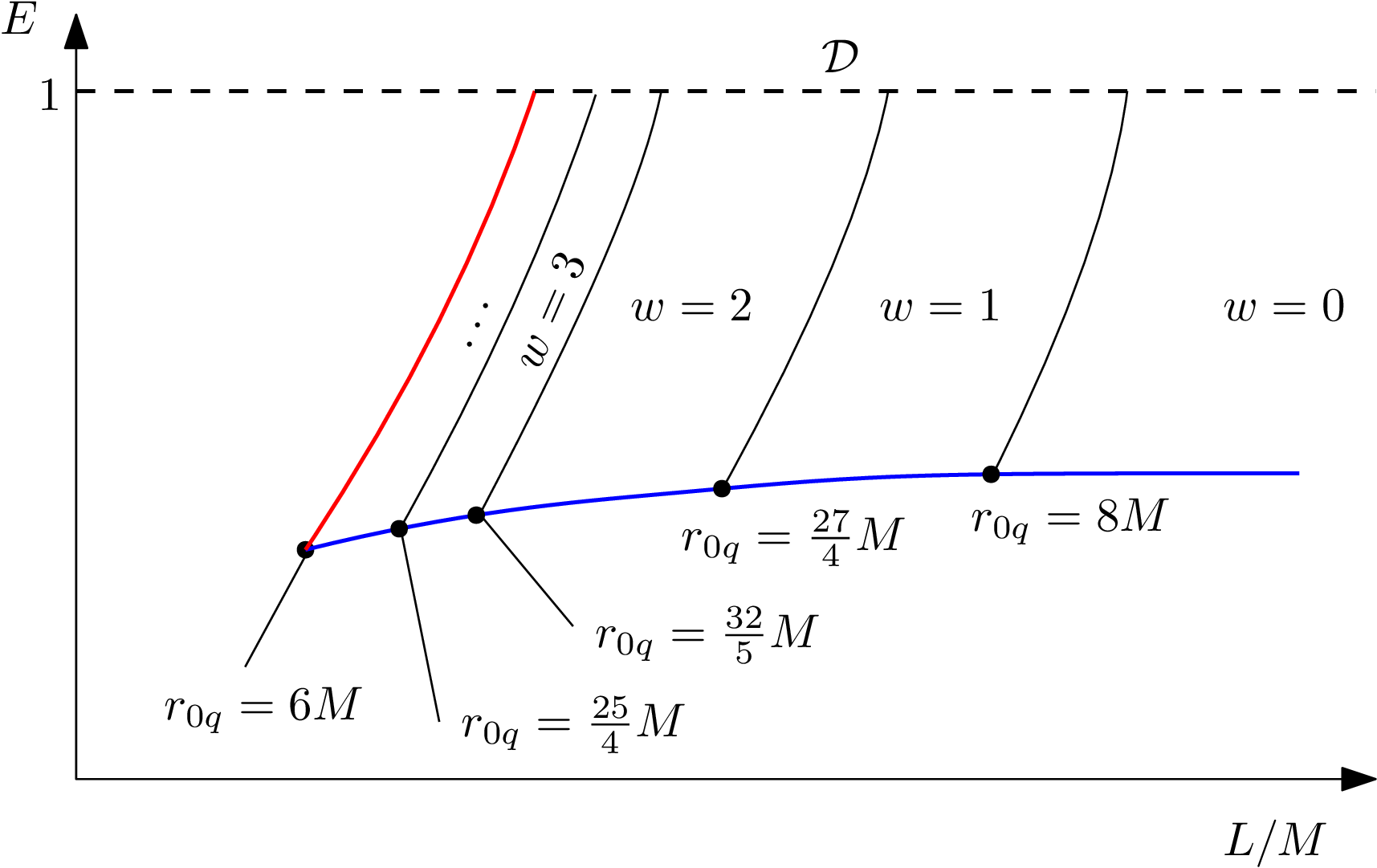}
 \caption{(Not to scale.) A sketch of the subdivision of the domain $\mathcal{D}$ of periodic bound orbits according to whirl number $w$. Note that this sketch is not to scale, as otherwise the domains with $w\geq2$ onward would be too narrow to be seen.}
 \label{fig_wdomains}
\end{figure}

\textbf{Subdivision of $\mathcal{D}$ according to whirl number.} 
We can now see how $\mathcal{D}$ can be partitioned according to whirl number $w$. From Eq.~\Eqref{perturb_q}, we have $w+\frac{v}{z}=\sqrt{\frac{r_{0q}}{r_0-6M}}-1$. Since $0\leq\frac{v}{z}<1$, we have $w\leq\sqrt{\frac{r_{0q}}{r_{0q}-6M}}-1< w+1$, or
\begin{align}
 \frac{6M(w+2)^2}{(w+1)(w+3)}< r_{0q} \leq\frac{6M(w+1)^2}{w(w+2)}. \label{w_domain}
\end{align}
So all periodic orbits of a given $w$ are branches which emanate from the stable circular branch at the domain specified by Eq.~\Eqref{w_domain}. The ranges for the first few $w$ are shown here:
\begin{align*}
 w=0:\quad 8M<&\;r_{0q}<\infty,\\
 w=1:\quad \frac{27}{4}M<&\;r_{0q}\leq8M,\\
 w=2:\quad \frac{32}{5}M<&\;r_{0q}\leq\frac{27}{4}M,\\
 w=3:\quad \frac{25}{4}M<&\;r_{0q}\leq\frac{32}{5}M,\\
            \vdots&
\end{align*}
The corresponding subdivisions of $\mathcal{D}$ are sketched in Fig.~\ref{fig_wdomains}.

Next, we observe that within each fixed $w$, the orbit $(1,w,0)$ is 
\begin{align}
 r_{0q}=\frac{6M(w+1)^2}{w(w+2)}, \label{z1_boundary}
\end{align}
which is the upper bound of \Eqref{w_domain}. In other words, within each $w$, the orbit with the largest $r_{0q}$ is the single-petal orbit $z=1$.

\textbf{Large $z$ limit at fixed $w$ and $v=1$.} For $z\geq 2$, and still with fixed $w$, we have the sequence 
\begin{align}
 r_0(2,w,v),\;r_0(3,w,v),\ldots,r_0(z,w,v),\ldots,
\end{align}
where for each $z$, we additionally have a finite subsequence where $v$ takes the possible values relatively prime to $z$. Let us first consider the case of fixed $v=1$. We find that the sequence $\{r_0(z,w,1)\}$ is increasing, since 
\begin{align}
 r_0(z+1,w,1)-r_0(z,w,1)=\frac{6M(1+2wz+2wz^2+2z^2+4z)}{(wz+w+1)(wz+w+3+2z)(wz+1)(wz+1+2z)},
\end{align}
which is positive. So the periodic orbit $(z+1,w,1)$ always lies to the right of $(z,w,1)$ in the $(L,E)$-plane. As $z$ is increased to infinity, the sequence of branches converges to the right boundary of its $w$ subdomain. On the other hand, as we have shown in Eq.~\Eqref{z1_boundary}, the boundary corresponds to the $z=1$ orbit. Therefore the infinite-$z$ limit of the sequence converges to $(1,w,0)$. We can verify this using Eq.~\Eqref{r0_zwv}: 
\begin{align}
 \lim_{z\rightarrow\infty}r_0(z,w,1)&=\lim_{z\rightarrow\infty}\frac{6M(z+v+wz)^2}{(wz+v)(wz+v+2z)}\nonumber\\
  &=\frac{6M(w+1)}{w(w+2)}\nonumber\\
  &=r_0(1,w,0).
\end{align}
Or, in terms of the branches themselves,
\begin{align}
 \lim_{z\rightarrow\infty}(z,w,1)=(1,w,0). \label{large_z}
\end{align}
For the case $w=0$, we find $\lim_{z\rightarrow\infty} r_0(z,0,1)=\infty$. However, recall that the infinite $r_0$ circular orbit coincides with $L\rightarrow\infty$ and $E\rightarrow1$, which is the non-relativistic limit. Therefore setting $w=0$ in Eq.~\Eqref{large_z} to write $\lim_{z\rightarrow\infty}(z,0,1)=(1,0,0)$ is consistent with the statement that $(1,0,0)$ is the non-relativistic Kepler ellipse.

\textbf{Large $w$ limit.} Returning to Eq.~\Eqref{r0_zwv}, we also have
\begin{align}
 \lim_{w\rightarrow\infty}r_0(z,w,v)=6M,
\end{align}
for any $z$ and $v$. This shows that the infinite whirl limit is the branch of unstable circular orbits. Intuitively, this can perhaps be understood as follows: Unstable circular orbits, when perturbed outwards,\footnote{Of course, inward perturbations of unstable circular orbits will typically send the particle plunging into the horizon.} will typically result in bound orbits with many whirls. These perturbed unstable orbits are the \emph{homoclinic orbits}. More details these orbits, including the more general case of Kerr spacetime, were studied in Refs.~\cite{Levin:2008yp,Mummery:2023hlo}. In terms of branches in the $(L,E)$-plane, this also suggests the interpretation that the branch of unstable circular orbit is itself a periodic orbit. More specifically, the limit $w\rightarrow\infty$ of the sequence of periodic orbits.

\begin{figure}
 \centering
 \includegraphics[scale=0.85]{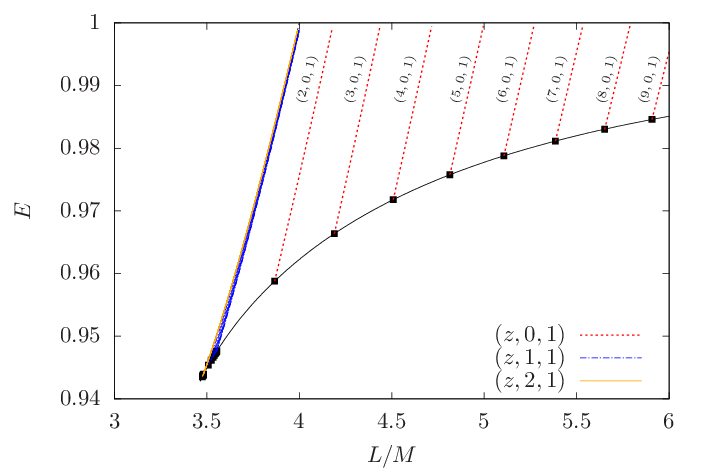}
 \includegraphics[scale=0.85]{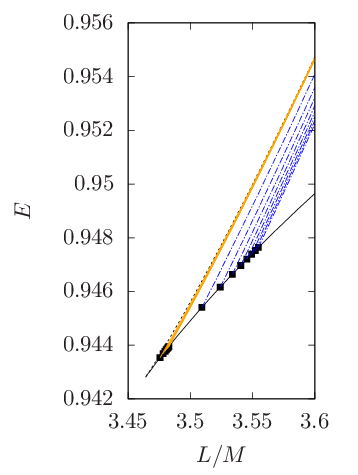}
 \caption{Angular momenta $L$ and energies $E$ of periodic orbits of $v=1$, $1\leq z\leq9$ with $w=0$ (dotted, red), $w=1$ (dash-dotted, blue), and $w=2$ (solid, orange). Here the horizontal axis is given in terms of dimensionless units $L/M$, where $M$ is the Schwarzschild mass parameter. The black squares shows the points where the periodic orbits meet the branch of stable circular orbits. For each $w$, the sequence of curves from left to right correspond to orbits of increasing $z$. As can be seen in the left panel, the orbits for $w=1$ and $w=2$ are closely packed near the curve of unstable circular orbits. The right panel shows the region near the ISCO in more detail.}
 \label{fig_ParamSpace}
\end{figure}

\textbf{Periodic orbits in the $(L,E)$-plane and the case $v>1$.} So far, we have inferred the distribution of orbits by studying the intersection points with the stable circular orbits at low eccentricity limits, primarily with the use of Eq.~\Eqref{r0_zwv}. We now verify the descriptions by obtaining the full structure of branches using the procedure discussed in Sec.~\ref{subsec_procedure} to obtain the precise values of $L$ and $E$ of the branches. We start with Fig.~\ref{fig_ParamSpace} where we plot the full branches of $v=0$ orbits.

As shown in Fig.~\ref{fig_ParamSpace}, the branches of periodic orbits can be partitioned according to the values of $w$. The orbits with $w=0$ occupy most of the space in the $(L,E)$-plane, with $w\geq1$ are tightly packed in the vicinity near the curve of unstable circular orbits. The resolution used in Fig.~\ref{fig_ParamSpace} is just barely able to show the $w=1$ and $w=2$ orbits. Going to higher $w>2$, the branches are packed even more closely together such that they are almost indistinguishible from the branch of unstable circular orbits when viewed at this scale. 

Finally, we turn to the orbits of $v>1$. According to Eq.~\Eqref{zv_constraint}, the set of possible $v$'s depend on $z$. To reiterate, unless $z=0$, the we have $1\leq v<z$ and $v$ must be relatively prime to $z$. Therefore it makes sense to view the various $v$'s as a subsequences depending on each given $z$. With this in mind, let us consider some fixed $w$ and $z$. The possible values of $v$ are the integers relatively prime to $z$. 

\begin{figure}
 \centering 
 \begin{subfigure}[b]{0.9\textwidth}
    \centering
    \includegraphics[scale=1]{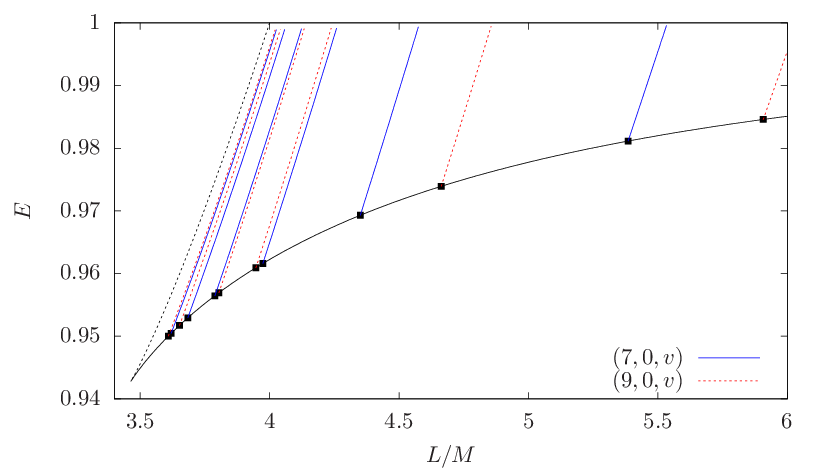}
    \caption{$w=0$.}
    \label{fig_vbandsw0}
  \end{subfigure}
  \begin{subfigure}[b]{0.9\textwidth}
    \centering
    \includegraphics[scale=1]{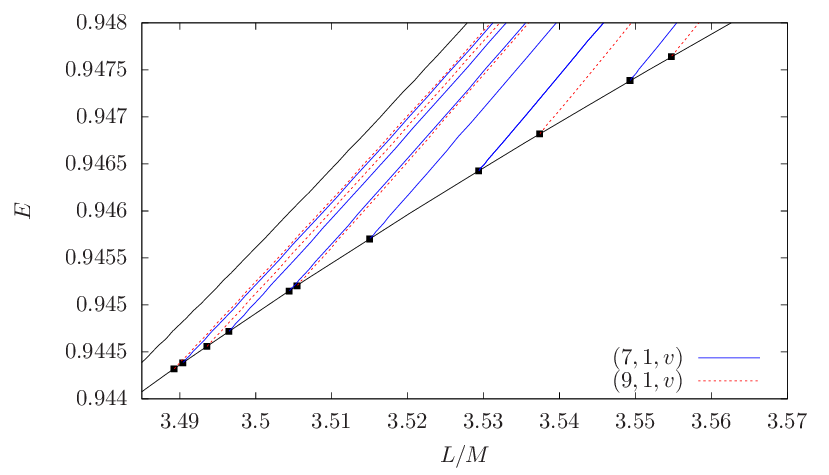}
    \caption{$w=1$.}
    \label{fig_vbandsw1}
  \end{subfigure}
 \caption{Periodic orbits for the case $w=0$ (a) and $w=1$ (b). The solid blue curves are for $(7,w,v)$, where the sequence of increasing $v=1$, $2$, $3$, $4$, $5$ and $6$ are sequence the branches going from right to left. (That is, the larger the $v$, at fixed $z$ and $w$, the smaller the angular momentum.) Similarly, the dotted red curves are for $(9,w,v)$ where the sequence $v=1$, $2$, $4$, $5$, $7$, and $8$ go from right to left.}
 \label{fig_vbands}
\end{figure}

We can show using Eq.~\Eqref{r0_zwv}, the sequence $r_0(z,w,v)$ is a decreasing (finite) sequence. Let $r_0(z,w,v)$ be a particular term of the sequence, and the next term is $r_0(z,w,v+k)$ for an appropriate $k$. Then 
\begin{align}
 r_0(z,w,v+k)-r_0(z,w,v)=-\frac{6Mz^2k(k+2z+2wz+2v)}{(wz+v+k)(wz+v+k+2z)(wz+v)(wz+v+2z)},
\end{align}
which is negative. So the subsequent branch $(z,w,v+k)$ always lies to the left of $(z,w,v)$. Of course, if $z$ is large, the subsequence $(z,w,v)$ will have more branches. In Fig.~\ref{fig_vbands}, we show the subsequence of $v$'s for $z=7$ and $z=9$, for $w=0$ and $w=1$, respectively. We observe that as $v$ increases, the branches quickly approach the left boundary of their respective $w$ subdomains.

\begin{figure}
 \begin{subfigure}[b]{0.49\textwidth}
    \centering
    \includegraphics[scale=0.8]{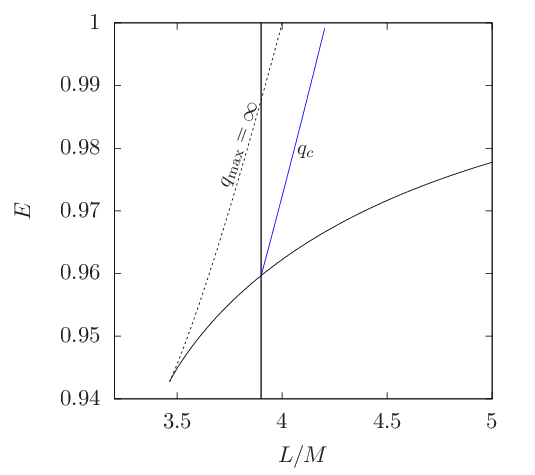}
    \caption{$L=3.9M$.}
    \label{fig_qcqmax1}
  \end{subfigure}
  \begin{subfigure}[b]{0.49\textwidth}
    \centering
    \includegraphics[scale=0.8]{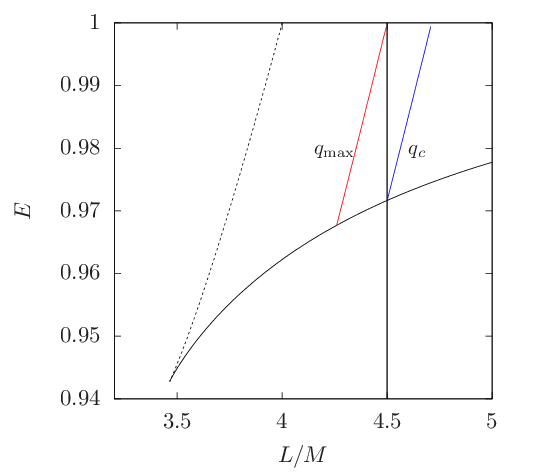}
    \caption{$L=4.5M$.}
    \label{fig_qcqmax2}
  \end{subfigure}
  \caption{Intersection of orbits of various $q$ with (a) $L=3.9M$ and (b) $L=4.5M$. In (a),  $q_c=0.475387316$ and the constant-$L$ line intersects all branches in the range $q_c\leq q\leq\infty$. For (b), $q_c=0.251678434$ and $q_{\mathrm{max}}=0.310993859$ and the constant-$L$ line intersects the branches in the range $q_c\leq q\leq q_{\mathrm{max}}$.}
  \label{fig_qcqmax}
\end{figure}

\textbf{Orbits with fixed $L$.} We now investigate the set of allowed orbits for fixed $L$. This corresponds to a vertical line in the $(L,E)$-plane. Such a line will intersect various $q$-branches. Let us denote by $q_c$ the branch of smallest $q$ which intersects this line, and $q_{\mathrm{max}}$ the branch with largest $q$. In other words, the vertical line of some fixed $L$, intersects the branches whose $q$-values lie in
\begin{align*}
 q_c\leq q\leq q_{\mathrm{max}}.
\end{align*}
This is Eq.~(12) in \cite{Levin:2008mq}. Since each rational $q$ correspond to a periodic branch which are curves of positive slope emanating from the stable circular orbit branch in $(L,E)$-space, one can directly see either from Eq.~\Eqref{q_max} or Fig.~\ref{fig_qcqmax} that the set of allowed orbits depend on whether $L$ is greater or less than $4M$. In particular,
\begin{align}
 q_c\leq q\leq \infty,\quad&\mbox{for}\quad 2\sqrt{3}M<L<4M,\nonumber\\
 q_c\leq q\leq q_{\mathrm{max}}<\infty,\quad&\mbox{for}\quad L>4M.
\end{align}

In closing this subsection, we provide a summary of the discussions so far: We can see that the domain $\mathcal{D}$ of bound orbits in the $(L,E)$-space can be partitioned into sub-domains for each $w$, as sketched in Fig.~\ref{fig_wdomains}. The sequence of sub-domains $\mathcal{D}_w\subset\mathcal{D}$ is an infinite sequence, as $w\rightarrow\infty$. The zero-whirl orbits $\mathcal{D}_0$ is the largest domain and contains the non-relativistic limit $L/M\rightarrow\infty$ and $E\rightarrow1$. Whereas $\mathcal{D}_{w>1}$ is a sequence of decreasing size. Each domain can be seen has having a left and right boundary. Within each $\mathcal{D}_w$, the right boundary is the limit $\lim_{z\rightarrow\infty}(z,w,v)=(1,w,0)$, coinciding with the left boundary of $\mathcal{D}_{w-1}$. It is worth noting that orbits with non-zero whirl all lie within a very narrow band in the $(L,E)$-plane. In that sense, orbits with $w\geq1$ only occur for a very narrow range of $L$ and $E$. The limit $\mathcal{D}_{w\rightarrow\infty}$ converges to the branch of unstable circular orbits.

\subsection{Stable vs unstable circular orbits} 

The observations so far reveal the different roles played by the stable and unstable circular orbits in the context of Levin and Perez-Giz's scheme. Outside of this scheme, the straightforward way to understand circular orbits is simply via Eq.~\Eqref{circular_EL}: they can be plotted as a curve in the $(L,E)$-plane parametrised by $r_0$, as was done in Fig.~\ref{fig_circular}. This curve has a cusp at $r_0=6M$ representing the ISCO. The cusp separates the curve into two parts, namely the stable ($r_0>6M$) and unstable ($3M<r_0<6M$) circular orbits.

However, the discussions in Sec.~\ref{subsec_stableparam} shows all branches of bound orbits terminate at the branch of stable circular orbits, and that the $w\rightarrow\infty$ limit converges to the unstable circular orbits. So rather than viewing $r_0>6M$ and $3M<r_0<6M$ as two parts of a single curve, in the present context it is appropriate to view only $r_0>6M$ as the `main branch' parametrising all branches of bound orbits. In other words, the set of all periodic orbits are captured by the blue segment of Fig.~\ref{fig_wdomains}, because arbitrarily close to any point on this main branch, there is a branch of periodic orbit that emanates from it. The red segment of Fig.~\ref{fig_wdomains} of unstable circular orbit can be viewed the last branch emanating from the end of the `main' (blue) branch. In other words, it seems more appropriate to view the unstable circular orbits part of the set of periodic orbits.

To reiterate previous discussions, all bound orbits can be associated with a number $0<q<\infty$, and every $q$ corresponds to a branch that emanates from some point on stable circular orbit branch and ends at the line $E=1$. The unstable circular orbit is simply the branch (loosely speaking\footnote{It is perhaps more precise to describe this branch as the $q\rightarrow\infty$ limit of the sequence of branches.}) `$q=\infty$' that emanates from $r_0=6M$.

\section{Conclusion} \label{sec_conclusion}

We have revisited the problem of periodic orbits in Schwarzschild spacetime. The geodesic equations for this spacetime can be solved by compact expressions in terms of the elliptic integrals of the first kind. By expressing the parameters of the equations of motion in terms of the eccentricity $e$ and the latus rectum $\lambda$, one can quickly identify the energy and angular momentum of any periodic orbits of a specified $(z,w,v)$ and $e$.

With this procedure, the set of all periodic orbits can be charted in the $(L,E)$-plane, where $L$ is the angular momentum and $E$ is the energy. Each periodic orbit consists positive-slope curves parametrised by $0\leq e<1$. The point $e=0$ is where the branch meets the branch of stable circular orbits. Running upward (increasing $E$) along the curve corresponds to increasing $e$, until the limit $e\rightarrow1$ where it meets the line $E=1$, which is the limit of escaping trajectories. 

The domain $\mathcal{D}$ of bound orbits in the $(L,E)$-plane is `foliated' by the branches of periodic orbits, and can be partitioned according to whirl numbers $w$. The domain $w=0$ contains the non-relativistic Kepler limit, and increasing $w$ corresponds to the sequence of domains of decreasing area, which converges to the branch of unstable circular orbits in the limit $w\rightarrow\infty$. Each domain has a `left' and `right' boundary. In particular, the right boundary is the limit $\lim_{z\rightarrow\infty}(z,w,v)=(1,w,0)$, and also it converges to the left boundary of the adjacent domain. We find that orbits of non-zero whirl lie in a very narrow band in the $(L,E)$-plane, suggesting that these kinds of orbits are relatively less probable for a generically chosen $L$ and $E$ for a particle orbit.

\section*{Acknowledgments}
 
Y.-K.~L is supported by Xiamen University Malaysia Research Fund (Grant No. XMUMRF/ 2021-C8/IPHY/0001).

\appendix 

\section{The Levin--Perez-Giz taxonomy} \label{app_taxonomy}

\begin{figure}
 \begin{subfigure}[b]{0.245\textwidth}
    \centering
    \includegraphics[scale=0.5]{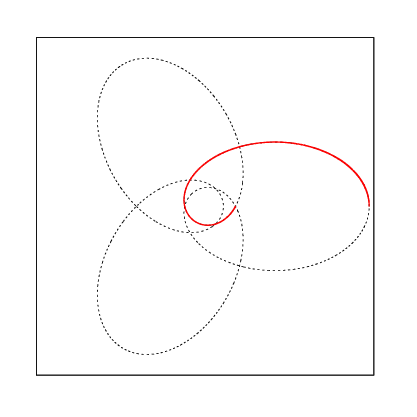}
    \caption{$\Delta\phi=2\pi$.}
    \label{fig_exposition301a}
  \end{subfigure}
  \begin{subfigure}[b]{0.245\textwidth}
    \centering
    \includegraphics[scale=0.5]{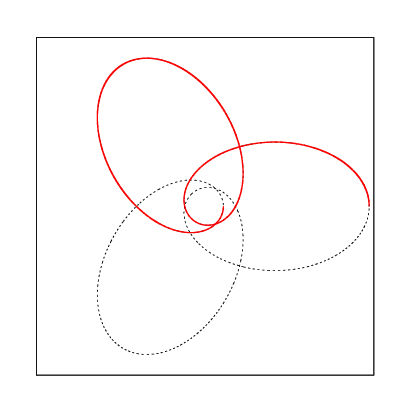}
    \caption{$\Delta\phi=4\pi$.}
    \label{fig_exposition301b}
  \end{subfigure}
  \begin{subfigure}[b]{0.245\textwidth}
    \centering
    \includegraphics[scale=0.5]{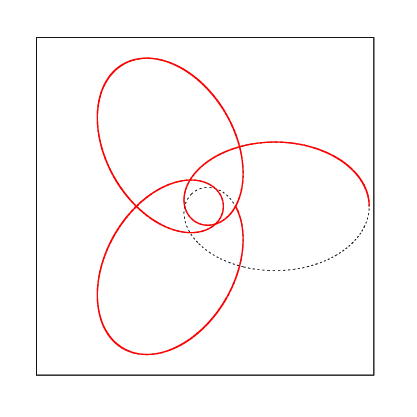}
    \caption{$\Delta\phi=6\pi$.}
    \label{fig_exposition301c}
  \end{subfigure}
  \begin{subfigure}[b]{0.245\textwidth}
    \centering
    \includegraphics[scale=0.5]{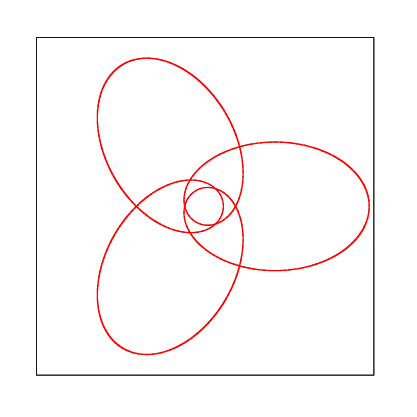}
    \caption{$\Delta\phi=8\pi$.}
    \label{fig_exposition301d}
  \end{subfigure}
  \caption{The trajectory of the $(3,0,1)$ orbit after $\phi$ has evolved by $2\pi$, $4\pi$, $6\pi$, and $8\pi$, respectively.}
  \label{fig_exposition301}
\end{figure}

\begin{figure}
 \begin{subfigure}[b]{0.33\textwidth}
    \centering
    \includegraphics[scale=0.5]{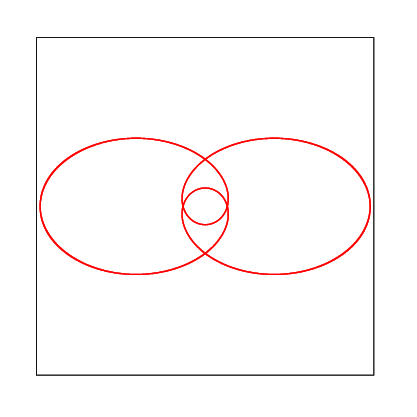}
    \caption{$(2,0,1)$.}
    \label{fig_otherz_201}
  \end{subfigure}
  \begin{subfigure}[b]{0.33\textwidth}
    \centering
    \includegraphics[scale=0.5]{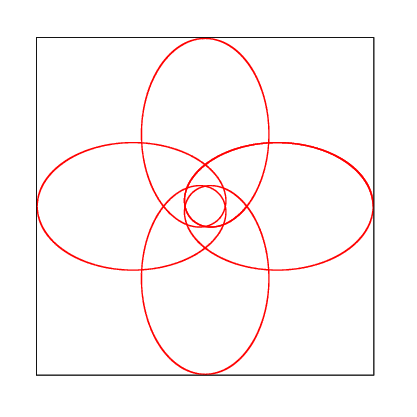}
    \caption{$(4,0,1)$.}
    \label{fig_otherz_401}
  \end{subfigure}
  \begin{subfigure}[b]{0.33\textwidth}
    \centering
    \includegraphics[scale=0.5]{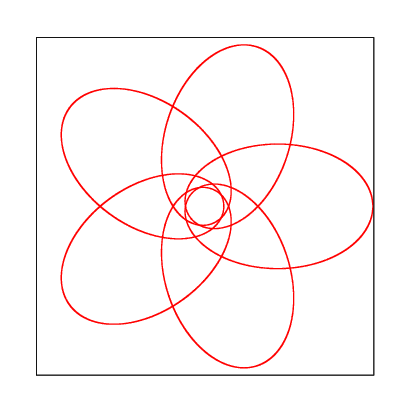}
    \caption{$(5,0,1)$.}
    \label{fig_otherz_501}
  \end{subfigure}
  \caption{Examples of $z=2$, $z=4$, and $z=5$ for orbits $(z,0,1)$.}
  \label{fig_otherz}
\end{figure}

According to Levin and Perez-Giz's taxonomy scheme \cite{Levin:2008mq}, rational orbits occur for
\begin{align}
 q=w+\frac{v}{z},
\end{align}
for positive integers $(z,w,v)$ where
\begin{align}
 1\leq& v\leq z-1,\quad \mbox{if }z>1,\quad \mbox{$z$ and $v$ co-prime},\nonumber\\
 v&=0,\quad \mbox{if } z=1.
\end{align}
We briefly review the meanings of the integers $z$, $v$, and $w$ in the following. First, $z$ is the number of `leaves' or `petals' in the orbit, called \emph{zoom} by Levin and Perez-Giz. They are the distinctive lobes which become elongated as $e$ approaches $1$. The leaves are uniformly distributed in angle. Hence for an orbit with $z$ leaves, each leaves are separated by an angle of $2\pi/z$. Fig.~\ref{fig_exposition301} shows an example of a $z=3$ (`three-leaf') orbit. Some examples for other values of $z$ are shown in Fig.~\ref{fig_otherz}.

The integer $v$ determines how the particle traces out each $z>1$ leaves for a given $z$. Now, each leaf is located at angular positions $2\pi k/z$ for $k=0,\ldots,z-1$. The integer $v$ is the how the particles skips each leaf in the sequence
\begin{align*}
 0,\frac{2\pi}{z},\;\frac{2\pi(2)}{z},\ldots,\frac{2\pi(z-1)}{z}.
\end{align*}

\begin{figure}
\centering
 \begin{subfigure}[b]{0.245\textwidth}
    \centering
    \includegraphics[scale=0.5]{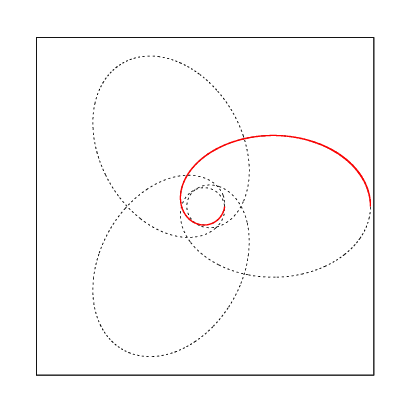}
    \caption{$\Delta\phi=2\pi$.}
    \label{fig_exposition302a}
  \end{subfigure}
  \begin{subfigure}[b]{0.245\textwidth}
    \centering
    \includegraphics[scale=0.5]{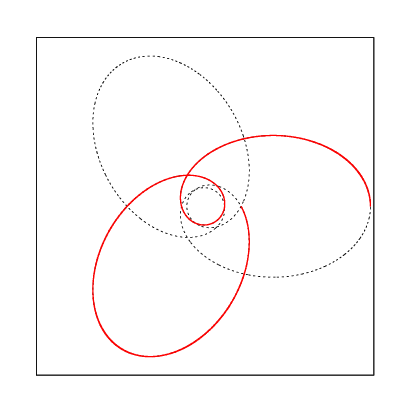}
    \caption{$\Delta\phi=4\pi$.}
    \label{fig_exposition302b}
  \end{subfigure}
  \begin{subfigure}[b]{0.245\textwidth}
    \centering
    \includegraphics[scale=0.5]{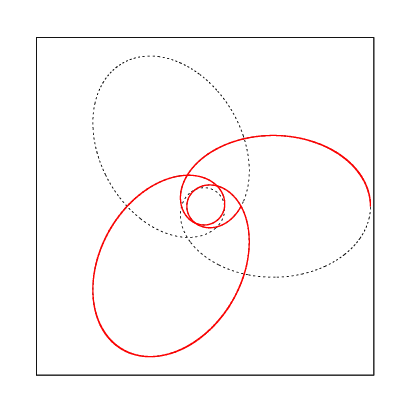}
    \caption{$\Delta\phi=6\pi$.}
    \label{fig_exposition302c}
  \end{subfigure}
  \begin{subfigure}[b]{0.245\textwidth}
    \centering
    \includegraphics[scale=0.5]{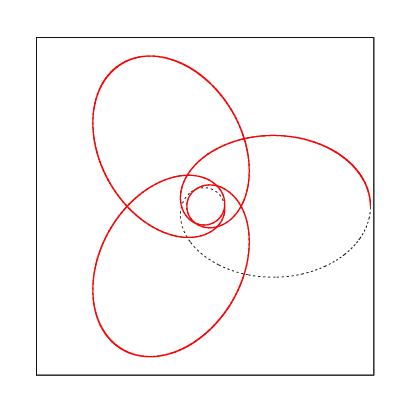}
    \caption{$\Delta\phi=8\pi$.}
    \label{fig_exposition302d}
  \end{subfigure}
  \begin{subfigure}[b]{0.245\textwidth}
    \centering
    \includegraphics[scale=0.5]{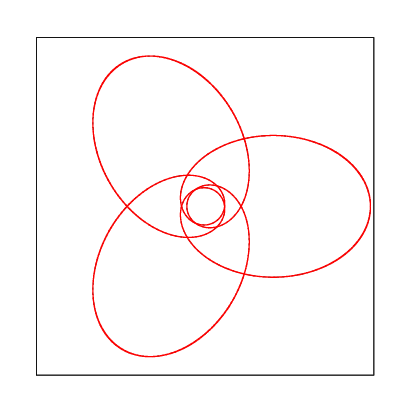}
    \caption{$\Delta\phi=10\pi$.}
    \label{fig_exposition302e}
  \end{subfigure}
  \caption{The trajectory of the $(3,0,2)$ orbit after $\phi$ has evolved by $2\pi$, $4\pi$, $6\pi$, $8\pi$, and $10\pi$, respectively.}
  \label{fig_exposition302}
\end{figure}

Note that $v$ must be relatively prime to $z$ otherwise there exist a leaf (or leaves) which are always skipped by the particle, contradicting the fact that the orbit consists of $z$ leaves. In Fig.~\ref{fig_exposition302}, we show an example of an orbit with $z=3$ and $v=2$. For $z=3$, the sequence of leaves are located at angles
\begin{align*}
 0,\frac{2\pi}{3},\;\frac{4\pi}{3}.
\end{align*}
We have fixed initial conditions so that the particle starts at the tip of the 0-leaf. Here, $v=2$ means it skips one leaf in the sequence (namely $\frac{2\pi}{3}$) and proceeds to $\frac{4\pi}{3}$. Then skips one leaf again (namely 0) and goes out to $\frac{2\pi}{3}$.

\begin{figure}
 \centering
 \begin{subfigure}[b]{0.245\textwidth}
    \centering
    \includegraphics[scale=0.5]{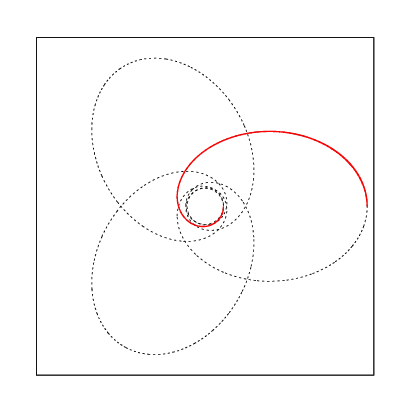}
    \caption{$\Delta\phi=2\pi$.}
    \label{fig_exposition311a}
  \end{subfigure}
  \begin{subfigure}[b]{0.245\textwidth}
    \centering
    \includegraphics[scale=0.5]{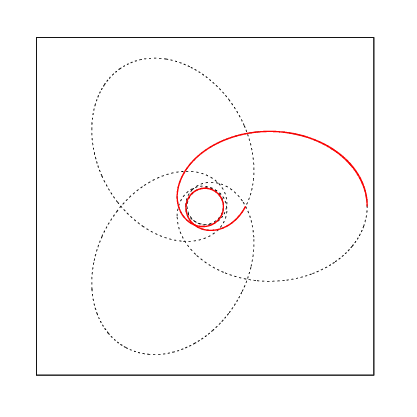}
    \caption{$\Delta\phi=4\pi$.}
    \label{fig_exposition311b}
  \end{subfigure}
  \begin{subfigure}[b]{0.245\textwidth}
    \centering
    \includegraphics[scale=0.5]{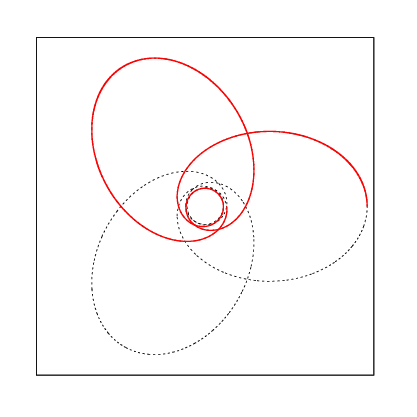}
    \caption{$\Delta\phi=6\pi$.}
    \label{fig_exposition311c}
  \end{subfigure}
  \begin{subfigure}[b]{0.245\textwidth}
    \centering
    \includegraphics[scale=0.5]{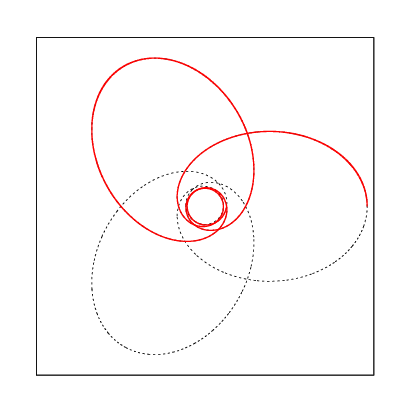}
    \caption{$\Delta\phi=8\pi$.}
    \label{fig_exposition311d}
  \end{subfigure}
  \begin{subfigure}[b]{0.245\textwidth}
    \centering
    \includegraphics[scale=0.5]{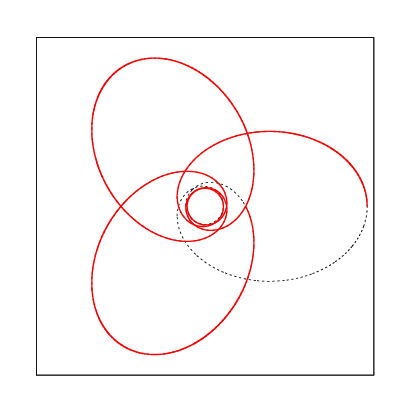}
    \caption{$\Delta\phi=10\pi$.}
    \label{fig_exposition311e}
  \end{subfigure}
  \begin{subfigure}[b]{0.245\textwidth}
    \centering
    \includegraphics[scale=0.5]{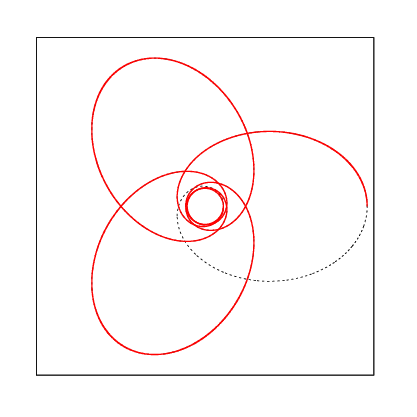}
    \caption{$\Delta\phi=12\pi$.}
    \label{fig_exposition311f}
  \end{subfigure}
  \begin{subfigure}[b]{0.245\textwidth}
    \centering
    \includegraphics[scale=0.5]{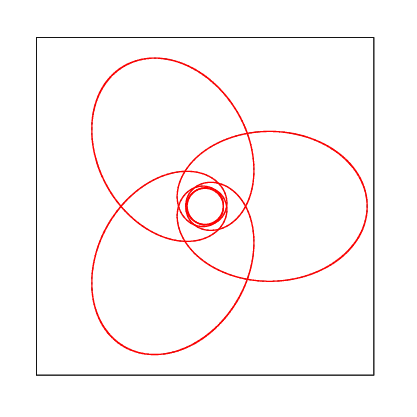}
    \caption{$\Delta\phi=14\pi$.}
    \label{fig_exposition311g}
  \end{subfigure}
  \caption{The trajectory of the $(3,1,1)$ ($z=3$ leaf) orbit after $\phi$ has evolved by $2\pi$, $4\pi$, $6\pi$, $8\pi$, $10\pi$, $12\pi$, and $14\pi$ respectively. For $w=1$, we see that the particle completes one additional inner round before proceeding out to the next leaf of the trajectory. In general, the particle executes $w$ additional inner circuits before proceeding to the next leaf for any non-negative integer $w$.}
  \label{fig_exposition311}
\end{figure}

Finally, the integer $w$ is the \emph{whirl} number. In between successive leaves (for any $v$), the particle spends some time near the minimum radius. Typically the angular velocity will be relatively large, and it may execute multiple circuits of $\phi$ before proceeding to the next leaf. The number of circuits is $w$. In Fig.~\ref{fig_exposition311}, we show an example of a $(3,1,1)$ orbit. Particularly $w=1$ and so in between successive leaves, we see that the particle executes $w=1$ extra circuit around the black hole before going to the next leaf.

\bibliographystyle{period}

\bibliography{period}

\end{document}